\documentstyle[twoside]{article}

\catcode`\@=11
\long\def\@makefntext#1{
\protect\noindent \hbox to 3.2pt {\hskip-.9pt  
$^{{\eightrm\@thefnmark}}$\hfil}#1\hfill}		

\def\thefootnote{\fnsymbol{footnote}}
\def\@makefnmark{\hbox to 0pt{$^{\@thefnmark}$\hss}}	
	
\def\ps@myheadings{\let\@mkboth\@gobbletwo
\def\@oddhead{\hbox{}
\rightmark\hfil\eightrm\thepage}   
\def\@oddfoot{}\def\@evenhead{\eightrm\thepage\hfil
\leftmark\hbox{}}\def\@evenfoot{}
\def\sectionmark##1{}\def\subsectionmark##1{}}

\oddsidemargin=\evensidemargin
\renewcommand{\thefootnote}{\fnsymbol{footnote}}

\newcounter{sectionc}\newcounter{subsectionc}\newcounter{subsubsectionc}
\renewcommand{\section}[1] {\vspace{12pt}\addtocounter{sectionc}{1} 
\setcounter{subsectionc}{0}\setcounter{subsubsectionc}{0}\noindent 
	{\tenbf\thesectionc. #1}\par\vspace{5pt}}
\renewcommand{\subsection}[1] {\vspace{12pt}\addtocounter{subsectionc}{1} 
	\setcounter{subsubsectionc}{0}\noindent 
	{\bf\thesectionc.\thesubsectionc. {\kern1pt \bfit #1}}\par\vspace{5pt}}
\renewcommand{\subsubsection}[1] {\vspace{12pt}\addtocounter{subsubsectionc}{1}
	\noindent{\tenrm\thesectionc.\thesubsectionc.\thesubsubsectionc.
	{\kern1pt \tenit #1}}\par\vspace{5pt}}
\newcommand{\nonumsection}[1] {\vspace{12pt}\noindent{\tenbf #1}
	\par\vspace{5pt}}

\newcounter{appendixc}
\newcounter{subappendixc}[appendixc]
\newcounter{subsubappendixc}[subappendixc]
\renewcommand{\thesubappendixc}{\Alph{appendixc}.\arabic{subappendixc}}
\renewcommand{\thesubsubappendixc}
	{\Alph{appendixc}.\arabic{subappendixc}.\arabic{subsubappendixc}}

\renewcommand{\appendix}[1] {\vspace{12pt}
        \refstepcounter{appendixc}
        \setcounter{figure}{0}
        \setcounter{table}{0}
        \setcounter{lemma}{0}
        \setcounter{theorem}{0}
        \setcounter{corollary}{0}
        \setcounter{definition}{0}
        \setcounter{equation}{0}
        \renewcommand{\thefigure}{\Alph{appendixc}.\arabic{figure}}
        \renewcommand{\thetable}{\Alph{appendixc}.\arabic{table}}
        \renewcommand{\theappendixc}{\Alph{appendixc}}
        \renewcommand{\thelemma}{\Alph{appendixc}.\arabic{lemma}}
        \renewcommand{\thetheorem}{\Alph{appendixc}.\arabic{theorem}}
        \renewcommand{\thedefinition}{\Alph{appendixc}.\arabic{definition}}
        \renewcommand{\thecorollary}{\Alph{appendixc}.\arabic{corollary}}
        \renewcommand{\theequation}{\Alph{appendixc}.\arabic{equation}}
        \noindent{\tenbf Appendix \theappendixc #1}\par\vspace{5pt}}
\newcommand{\subappendix}[1] {\vspace{12pt}
        \refstepcounter{subappendixc}
        \noindent{\bf Appendix \thesubappendixc. {\kern1pt \bfit #1}}
	\par\vspace{5pt}}
\newcommand{\subsubappendix}[1] {\vspace{12pt}
        \refstepcounter{subsubappendixc}
        \noindent{\rm Appendix \thesubsubappendixc. {\kern1pt \tenit #1}}
	\par\vspace{5pt}}

\newcommand{\textlineskip}{\baselineskip=13pt}
\newcommand{\smalllineskip}{\baselineskip=10pt}

\def\eightcirc{
\begin{picture}(0,0)
\put(4.4,1.8){\circle{6.5}}
\end{picture}}
\def\eightcopyright{\eightcirc\kern2.7pt\hbox{\eightrm c}} 

\newcommand{\copyrightheading}[1]
	{\vspace*{-2.5cm}\smalllineskip{\flushleft
	{\footnotesize International Journal of Modern Physics B, #1}\\
	{\footnotesize $\eightcopyright$\, World Scientific Publishing
	 Company}\\
	 }}

\newcommand{\publisher}[2]{{\begin{center}\footnotesize\smalllineskip 
	Received #1\\
	Revised #2
	\end{center}
	}}

\def\abstracts#1#2#3{{
	\centering{\begin{minipage}{4.5in}\baselineskip=10pt\footnotesize
	\parindent=0pt #1\par 
	\parindent=15pt #2\par
	\parindent=15pt #3
	\end{minipage}}\par}} 

\renewenvironment{thebibliography}[1]			
	{\frenchspacing
	 \ninerm\baselineskip=11pt
	 \begin{list}{\arabic{enumi}.}
	{\usecounter{enumi}\setlength{\parsep}{0pt}
	 \setlength{\leftmargin 12.7pt}{\rightmargin 0pt} 
	 \setlength{\itemsep}{0pt} \settowidth
	{\labelwidth}{#1.}\sloppy}}{\end{list}}

\newcounter{itemlistc}
\newcounter{romanlistc}
\newcounter{alphlistc}
\newcounter{arabiclistc}

\newcommand{\fcaption}[1]{
        \refstepcounter{figure}
        \setbox\@tempboxa = \hbox{\footnotesize Fig.~\thefigure. #1}
        \ifdim \wd\@tempboxa > 5in
           {\begin{center}
        \parbox{5in}{\footnotesize\smalllineskip Fig.~\thefigure. #1}
            \end{center}}
        \else
             {\begin{center}
             {\footnotesize Fig.~\thefigure. #1}
              \end{center}}
        \fi}

\newcommand{\tcaption}[1]{
        \refstepcounter{table}
        \setbox\@tempboxa = \hbox{\footnotesize Table~\thetable. #1}
        \ifdim \wd\@tempboxa > 5in
           {\begin{center}
        \parbox{5in}{\footnotesize\smalllineskip Table~\thetable. #1}
            \end{center}}
        \else
             {\begin{center}
             {\footnotesize Table~\thetable. #1}
              \end{center}}
        \fi}

\def\@citex[#1]#2{\if@filesw\immediate\write\@auxout
	{\string\citation{#2}}\fi
\def\@citea{}\@cite{\@for\@citeb:=#2\do
	{\@citea\def\@citea{,}\@ifundefined
	{b@\@citeb}{{\bf ?}\@warning
	{Citation `\@citeb' on page \thepage \space undefined}}
	{\csname b@\@citeb\endcsname}}}{#1}}

\newif\if@cghi
\def\cite{\@cghitrue\@ifnextchar [{\@tempswatrue
	\@citex}{\@tempswafalse\@citex[]}}
\def\citelow{\@cghifalse\@ifnextchar [{\@tempswatrue
	\@citex}{\@tempswafalse\@citex[]}}
\def\@cite#1#2{{$\null^{#1}$\if@tempswa\typeout
	{IJCGA warning: optional citation argument 
	ignored: `#2'} \fi}}

\def\pmb#1{\setbox0=\hbox{#1}
	\kern-.025em\copy0\kern-\wd0
	\kern.05em\copy0\kern-\wd0
	\kern-.025em\raise.0433em\box0}

\def\fnt#1#2{\footnotetext{\kern-.3em
	{$^{\mbox{\scriptsize #1}}$}{#2}}}

\def\fpage#1{\begingroup
\voffset=.3in
\thispagestyle{empty}\begin{table}[b]\centerline{\footnotesize #1}
	\end{table}\endgroup}

\def\runninghead#1#2{\pagestyle{myheadings}
\markboth{{\protect\footnotesize\it{\quad #1}}\hfill}
{\hfill{\protect\footnotesize\it{#2\quad}}}}
\font\tenrm=cmr10
\font\tenit=cmti10 
\font\tenbf=cmbx10
\font\bfit=cmbxti10 at 10pt
\font\ninerm=cmr9

\font\eightrm=cmr8

\def\qed{\hbox{${\vcenter{\vbox{			
   \hrule height 0.4pt\hbox{\vrule width 0.4pt height 6pt
   \kern5pt\vrule width 0.4pt}\hrule height 0.4pt}}}$}}

\renewcommand{\thefootnote}{\fnsymbol{footnote}}	

\def\bsc{{\sc a\kern-6.4pt\sc a\kern-6.4pt\sc a}}	
\def\bflatex{\bf L\kern-.30em\raise.3ex\hbox{\bsc}\kern-.14em 
T\kern-.1667em\lower.7ex\hbox{E}\kern-.125em X} 

\def\d{\delta}
\def\e{\epsilon}

\def\l{\lambda}

\def\t{\theta}

\def\D{\Delta}
\def\L{\Lambda}

\def\beq{\begin{equation}}
\def\eeq{\end{equation}}
\def\bea{\begin{eqnarray}}
\def\eea{\end{eqnarray}}
\def\ba{\begin{array}}
\def\ea{\end{array}}
\def\no{\nonumber}
\def\lt{\left}
\def\rt{\right}
\newcommand{\bq}{\begin{quote}}
\newcommand{\eq}{\end{quote}}

\begin{document}

\runninghead{R-matrices and the Tensor Product Graph Method}
{R-matrices and the Tensor Product Graph Method}
\normalsize\textlineskip
\thispagestyle{empty}
\setcounter{page}{1}

\copyrightheading{}			

\vspace*{0.88truein}

\fpage{1}
\centerline{\bf R-matrices and the Tensor Product Graph Method}
\vspace*{0.035truein}
\vspace*{0.37truein}
\centerline{\footnotesize Mark D. Gould}
\vspace*{0.015truein}
\centerline{\footnotesize\it Centre for Mathematical Physics, Department
of Mathematics, University of Queensland,}
\baselineskip=10pt
\centerline{\footnotesize\it Brisbane, Qld 4072, Australia}
\vspace*{10pt}
\centerline{\normalsize and}
\vspace*{10pt}
\centerline{\footnotesize Yao-Zhong Zhang}
\vspace*{0.015truein}
\centerline{\footnotesize\it Centre for Mathematical Physics, Department
of Mathematics, University of Queensland,}
\baselineskip=10pt
\centerline{\footnotesize\it Brisbane, Qld 4072, Australia}
\vspace*{0.225truein}
\publisher{(received date)}{(revised date)}

\vspace*{0.21truein}
\abstracts{A systematic method for constructing trigonometric R-matrices 
corresponding to the (multiplicity-free) tensor product of any two 
affinizable representations of a quantum algebra or superalgebra has been
developed by the Brisbane group and its collaborators. This method has been
referred to as the Tensor Product Graph Method. Here we describe 
applications of this method to untwisted and twisted quantum affine 
superalgebras.}{}{}



\vspace*{1pt}\textlineskip	
\section{Introduction}	        
\vspace*{-0.5pt}
\noindent
The (graded) Yang-Baxter equation (YBE) plays a central role in the
theory of (supersymmetric) quantum integrable systems. 
Solutions to the YBE are usually called R-matrices. The knowledge of
R-matrices has many physical applications. In one-dimensional lattice
models, R-matrices yield the Hamiltonians of quantum spin chains
\cite{Skl79}. In statistical mechanics, R-matrices define the Boltzmann
weights of exactly soluble models \cite{Bax82} and in integrable
quantum field theory they give rise to exact factorizable scattering
S-matrices \cite{Zam79}. So the construction of R-matrices 
is fundamental in the study of integrable systems.

Mathematical structures underlying the YBE and therefore R-matrices and
integrable models are quantum affine (super)algebras. A systematic
method for the construction of trigonometric R-matrices arising from
untwisted and twisted quantum affine (super)algebras has been developed
in \cite{Zha91,Del94,Del95,Del96,Gan96,Gou00} (see also \cite{Mac91} for
rational cases). This method is called the Tensor Product Graph (TPG)
method. The method enables one to construct spectral dependent
R-matrices corresponding to
the (multiplicity-free) tensor product of {\em any} two affinizable
representations of a quantum algebra or superalgebra.

In this contribution, we describe the TPG method in the context of
untwisted and twisted quantum affine superalgebras. Quantum superalgebras
are interesting since the tensor product decomposition often has
indecomposables and integrable models associated with them may in some
instances be interpreted as describing strongly correlated fermion systems
\cite{Ess92,Bra95}.


\textheight=7.8truein
\setcounter{footnote}{0}
\renewcommand{\thefootnote}{\alph{footnote}}

\section{Quantum Affine Superalgebras and Jimbo Equation}
\noindent
Let us first of all recall some facts about the affine superalgebra
${\cal G}^{(k)},~k=1,2$. Let ${\cal G}_0$ be the fixed point subalgebra under
the diagram automorphism of ${\cal G}$ of order $k$. 
In the case of $k=1$, we have ${\cal G}_0\equiv{\cal G}$.
For $k=2$ we may decompose ${\cal G}$ as ${\cal G}_0\oplus {\cal G}_1$,
where $[{\cal G}_0, {\cal G}_1]\subset {\cal G}_1$. Let
$\psi$ be the highest root of ${\cal G}_0\equiv {\cal G}$ for $k=1$ 
and $\t$ be the highest weight of the ${\cal G}_0$-representation 
${\cal G}_1$ for $k=2$.

Quantum affine superalgebras $U_q[{\cal G}^{(k)}]$ are $q$-deformations
of the universal enveloping algebras $U[{\cal G}^{(k)}]$ of 
${\cal G}^{(k)}$.
We shall not give the defining relations for $U_q[{\cal G}^{(k)}]$, 
but mention that the action of the coproduct on its 
generators $\{h_i,~e_i,~f_i,~0\leq i\leq r\}$ is given by
\bea
\D(h_i)&=&h_i\otimes 1+1\otimes h_i,\no\\
\D(e_i)&=&e_i\otimes q^{\frac{h_i}{2}}+q^{-\frac{h_i}{2}}\otimes
          e_i,~~~~
\D(f_i)=f_i\otimes q^{\frac{h_i}{2}}+q^{-\frac{h_i}{2}}\otimes
          f_i.
\eea
Define an automorphism $D_z$ of $U_q[{\cal G}^{(k)}]$ by
\beq
D_z(e_i)=z^{k\d_{i0}}e_i,~~~~~D_z(f_i)=z^{-k\d_{i0}}f_i,~~~~ D_z(h_i)=h_i.
\eeq
Given any two minimal irreducible representations 
$\pi_\l$ and $\pi_\mu$ of $U_q[{\cal G}_0]$
and their affinizations to irreducible representations of $U_q[{\cal G}^{(k)}]$,
we obtain a one-parameter family of representations 
$\D_{\l\mu}^z$ of $U_q[{\cal G}^{(k)}]$ on
$V(\l)\otimes V(\mu)$ defined by
\beq
\D_{\l\mu}^z(a)=\pi_\l\otimes \pi_\mu\left((D_z\otimes 1)
\D(a)\right),~~~\forall a\in U_q[{\cal G}^{(k)}],
\eeq
where $z$ is the spectral parameter. 
Let $R^{\l\mu}(z)$ be the spectral dependent R-matrices associated with
$\pi_\l$ and $\pi_\mu$, which satisfies the YBE.
Moreover it obeys the intertwining properties:
\beq\label{Jimbo}
R^{\l\mu}(z)\,\D_{\l\mu}^z(a)=(\D^T)_{\l\mu}^z(a)\,R^{\l\mu}(z)
\eeq
which, according to Jimbo \cite{Jim89},
uniquely determine $R^{\l\mu}(z)$ up to a scalar function of $z$. 
We normalize $R^{\l\mu}(z)$ such that
$\check{R}^{\l\mu}(z)\check{R}^{\mu\l}(z^{-1})=I$, where
$\check{R}^{\l\mu}(z)=P\,R^{\l\mu}(z)$ with
$P:V(\l)\otimes V(\mu)\rightarrow V(\mu)\otimes V(\l)$
the usual graded permutation operator.

In order for the equation (\ref{Jimbo}) to hold for 
all $a\in U_q[{\cal G}^{(k)}]$ it is sufficient that it holds for all
$a\in U_q(\hat{L}_0)$ and in addition for the extra generator
$e_0$.
The relation for $e_0$ reads explicitly
\bea
&&
\check{R}^{\l\mu}(z)
\left(z\,\pi_\l(e_0)\otimes\pi_\mu(q^{h_0/2})+
\pi_\l(q^{-h_0/2})\otimes \pi_\mu(e_0)\right)
\no\\
&&
~~~~~=\left(\pi_\mu(e_0)\otimes\pi_\l(q^{h_0/2})+
z\,\pi_\mu(q^{-h_0/2})\otimes\pi_\l(e_0)\right)
\check{R}^{\l\mu}(z).\label{jimbo-eq}
\eea
Eq.(\ref{jimbo-eq}) is the Jimbo equation for $U_q[{\cal G}^{(k)}]$.

\section{Solutions to Jimbo Equation and Tensor Product Graph Method}
\noindent
Let $V(\l)$ and $V(\mu)$ denote any two minimal irreducible representations of
$U_q[{\cal G}^{(k)}]$. Assume the tensor product
module $V(\l)\otimes V(\mu)$ is completely reducible into
irreducible $U_q[{\cal G}_0]$-modules as
\beq\label{dec}
V(\l)\otimes V(\mu)=\bigoplus_\nu V(\nu)
\eeq
and there are no multiplicities in this decomposition.
We denote by $P_\nu^{\l\mu}$  the projection operator of
$V(\l)\otimes V(\mu)$ onto $V(\nu)$ and set
${\bf P}^{\l\mu}_\nu=\check{R}^{\l\mu}(1)\,P^{\l\mu}_\nu=
P^{\mu\l}_\nu\,\check{R}^{\l\mu}(1)$.
We may thus write
\beq
\check{R}^{\l\mu}(z)=\sum_\nu\,\rho_\nu(z)\,
  {\bf P}^{\l\mu}_\nu,~~~
  \rho_\nu(1)=1.
\eeq
Following our previous approach \cite{Del94}, the coefficients 
$\rho_\nu(z)$ may be determined according to the recursion
relation
\beq\label{rec}
\rho_\nu(z)=\frac{q^{C(\nu)/2}+\e_\nu\e_{\nu'}z\,q^{C(\nu')/2}}
{z\,q^{C(\nu)/2}+\e_\nu\e_{\nu'}\,q^{C(\nu')/2}}
\rho_{\nu'}(z),
\eeq
which holds for any $\nu\neq\nu'$ for which
\beq\label{edge}
P^{\l\mu}_\nu\left(\pi_\l(e_0)\otimes\pi_\mu(q^{h_0/2})\right)
P^{\l\mu}_{\nu'}\neq 0.
\eeq
Here $C(\nu)$ is the eigenvalue of the universal Casimir element
of ${\cal G}_0$ on $V(\nu)$ and $\e_\nu$ denotes the parity
of $V(\nu)\subseteq V(\l)\otimes V(\mu)$.

We note that $e_0\otimes q^{h_0/2}$ transforms under the adjoint action
of $U_q[{\cal G}_0]$ as the lowest weight of ${\cal G}_0$-module $V(\psi)$
[resp. $V(\t)$] for $k=1$ (resp. $k=2$) (i.e. as the lowest component of
a tensor operator).  Throughout we adopt the notation
\beq
<a>_\pm=\frac{1\pm z\,q^a}{z\pm q^a},
\eeq
so that the relation (\ref{rec}) may be expressed as
\beq
\rho_\nu(z)=\left\langle\frac{C(\nu')-C(\nu)}{2}\right\rangle
_{\epsilon_\nu\epsilon_{\nu'}}\,
\rho_{\nu'}(z).
\eeq

To graphically encode the recursion relations between
different $\rho_\nu$ we introduce the {\bf Extended TPG} for
$U_q[{\cal G}^{(1)}]$ and {\bf Extended Twisted TPG}
for $U_q[{\cal G}^{(2)}]$.
\vskip.1in
\noindent{\bf Definition~1:}\label{tpg}
The {\bf Extended TPG} associated to the tensor product 
$V(\l)\otimes V(\mu)$ is a graph
whose vertices are the irreducible modules
$V(\nu)$ appearing in the decomposition (\ref{dec}) of
$V(\l)\otimes V(\mu)$. There is an edge between two vertices $V(\nu)$
and $V(\nu')$ iff
\beq
V(\nu')\subset V_{adj}\otimes V(\nu) ~~~\mbox{ and }~~
\epsilon(\nu)\epsilon(\nu')=-1.\label{adj}
\eeq
The condition (\ref{adj}) is a necessary
condition for (\ref{edge}) corresponding to $U_q[{\cal G}^{(1)}]$
to hold. 

\vskip.1in
\noindent{\bf Definition~2:} 
The {\bf Extended Twisted TPG} which has the same set of nodes as the 
twisted TPG but has an edge between two vertices $\nu\neq\nu'$ whenever
\beq\label{edge1}
V(\nu')\subseteq V(\theta)\otimes V(\nu)
\eeq
and
\beq\label{edge2}
\epsilon_\nu \epsilon_{\nu'}=
\left\{\begin{array}{l}
+1~~~\mbox{if }V(\nu)\mbox{ and }V(\nu')\mbox{ are in the same
irreducible representation of }{\cal G}\\
-1~~~\mbox{if }V(\nu)\mbox{ and }V(\nu')\mbox{ are in different
irreducible representations of }{\cal G}.\end{array}\right.
\eeq
The conditions (\ref{edge1}) and (\ref{edge2}) are necessary
conditions for (\ref{edge}) corresponding to $U_q[{\cal G}^{(2)}]$
to hold. 

We will impose a relation (\ref{rec}) for every edge in the
extended (twisted) TPG but we will be imposing too many relations in
general. These relations may be inconsistent and we are therefore not 
guaranteed a solution.  If however a solution to the recursion relations
exists, then it must give the
unique correct solution to the Jimbo's equation.

\section{Examples of R-matrices for $U_q[gl(m|n)^{(1)}]$\label{slmn-e}}
\noindent
Throughout we introduce $\{\e_i\}_{i=1}^m$ and $\{\d_j\}_{j=1}^n$
which satisfy $(\e_i,\e_j)=\d_{ij},~(\d_i,\d_j)=-\d_{ij}$ and
$(\e_i,\d_j)=0$. As is well known,
every irreducible represetation  of $U_q[gl(m|n)]$ provides also an 
irreducible representation for $U_q[gl(m|n)^{(1)}]$.
Here, as examples, we will construct the R-matrices corresponding
to the following tensor product: rank $a$ antisymmetric tensor
with rank $b$ antisymmetric tensor of the same type.
Without loss of generality, we assume $m\geq a\geq b$
and the antisymmetric tensors to be contravariant.
The tensor product decomposition is
\beq
V(\l_a)\otimes V(\l_b)=\bigoplus_c V(\L_c)
\eeq
where, when $a+b\leq m$,
\beq
\l_b=\sum_{i=1}^b\e_i\,,~~~~
\Lambda_c=\sum_{i=1}^{a+c}\e_i
  +\sum_{i=1}^{b-c}\e_i,~~~~~c=0,1,\cdots,b
\eeq
and when $a+b>m$,
\bea
&&\Lambda_c=\sum_{i=1}^{a+c}\e_i
  +\sum_{i=1}^{b-c}\e_i,~~~~~c=0,1,\cdots,m-a\no\\
&&\Lambda_c=\sum_{i=1}^m\e_i+\sum_{i=1}^{b-c}\e_i+(a+c-m)\d_1,
  ~~~~~c=m-a+1,\cdots,b
\eea
The corresponding TPG is
\beq
\unitlength=1mm
\linethickness{0.4pt}
\begin{picture}(112.60,7.60)(10,12)
\put(50.00,15.00){\circle*{5.20}}
\put(65.00,15.00){\circle*{5.20}}
\put(95.00,15.00){\circle*{5.20}}
\put(110.00,15.00){\circle*{5.20}}
\put(44.00,15.00){\makebox(0,0)[rc]{$V(\l_a)\otimes V(\l_b)~=$}}
\put(50.00,11.00){\makebox(0,0)[ct]{$\Lambda_0$}}
\put(65.00,11.00){\makebox(0,0)[ct]{$\Lambda_1$}}
\put(95.00,11.00){\makebox(0,0)[ct]{$\Lambda_{b-1}$}}
\put(110.00,11.00){\makebox(0,0)[ct]{$\Lambda_b$}}
\put(110.00,15.00){\line(-1,0){21.00}}
\put(50.00,15.00){\line(1,0){21.00}}
\put(80.00,15.00){\makebox(0,0)[cc]{$\cdots$}}
\end{picture}
\eeq
which is consistent; such is always the case when a graph is a tree (
i.e. contains no closed loops). From the graph we obtain
\beq
{\breve{R}}^{\l_a,\l_b}(x)=\sum_{c=0}^b\prod_{i=1}^c\left\langle
  2i+a-b\right\rangle_-\,{\bf P}
 ^{\l_a,\l_b}_{\Lambda_c}
\eeq
The $a=b=1$ case had been worked out before, which is known
to give rise to the Perk-Schultz model R-matrices \cite{Per81,Baz88}.

\section{Examples of R-matrices for $U_q[gl(n|n)^{(2)}]$}
\noindent
To begin with, we introduce the concept of minimal representations.
By minimal irreducible representations of ${\cal G}$, we mean those
irreducible representations which are also irreducible under
the fixed subalgebra ${\cal G}_0$. 
We can determine R-matrices for any tensor product
$V(\l_a)\otimes V(\l_b)$ of two minimal representations 
$V(\l_a),~V(\l_b)$ of $U_q[gl(m|n)^{(2)}]$, where
$V(\l_a)$ is also irredcible model under $U_q[osp(m|n)]$ with the
corresponding $U_q[osp(m|n)]$ highest weight 
$\l_a=(\dot{0}|a,\dot{0})$. Recall that
for our case ${\cal G}_0\equiv osp(m|n)$ and $\t=\d_1+\d_2$.
Below we shall illustrate the method for the interesting case of $a=b,~m=n>2$, 
where an indecomposable appears in the tensor product decomposition.

The decomposition of the tensor product of two minimal
irreducible representations of $U_q[osp(m|n)]$: \cite{Gou00}
\beq
V(\l_a)\otimes V(\l_b)
   =\bigoplus^a_{c=0}\bigoplus^c_{k=0}V(k, a+b-2c);\label{vv-decom}
\eeq
here and throughout $V(a,b)$ denotes an irreducible $U_q[osp(m|n)]$
module with highest weight
$\l_{a,b}=(\dot{0}|a+b,a,\dot{0})$.
Note that one can only get an indecomposable in (\ref{vv-decom}) when
$m=n>2$ and $a+b-2c=0$. Since $a\leq b,~ c\leq a$, this 
can only occur when $a=b$ and $c=a$.
In that case the $U_q[osp(m|n)]$-modules $V(k,0),~
k=0,1,$ will form an indecomposable. From now on we denote by $V$ this
indecomposable module, and write the
$U_q[osp(n|n)]$ module decomposition (\ref{vv-decom}) as
\beq
V(\l_a)\otimes V(\l_a)=\bigoplus_{\nu} V(\nu)\bigoplus V,\label{decom-indecom}
\eeq
where the sum on $\nu$ is over the irreducible highest weights.
Note that $V$ contains a unique
submodule $\bar{V}(\d_1+\d_2)$ which is maximal, indecomposable and
cyclically generated by a maximal vector of weight $\d_1+\d_2$ such
that $V/\bar{V}(\d_1+\d_2)\cong V(\dot{0}|\dot{0})$ (the trivial
$U_q[osp(n|n)]$-module). Moreover $V$ contains a unique irreducible
submodule $V(\dot{0}|\dot{0})\subset \bar{V}(\d_1+\d_2)$. The usual form
of Schur's lemma applies to $\bar{V}(\d_1+\d_2)$  and so the space
of $U_q[osp(n|n)]$ invariants in ${\rm End}(V)$ has dimension 2. It is
spanned by the identity operator $I$ together with an invariant $N$
(unique up to scalar multiples) satisfying
\beq
N\,V=V(\dot{0}|\dot{0})\subset\bar{V}(\d_1+\d_2),~~~~
N\,\bar{V}(\d_1+\d_2)=(0).\label{N-action}
\eeq
It follows that $N$ is nilpotent, i.e. $ N^2=0$.

We can show \cite{Gou00} that the minimal irreducible $U_q[osp(n|n)]$ modules,
$V(\l_a)$, with highest weight $\l_a$, are affinizable to carry
irreducible representations of $U_q[gl(n|n)^{(2)}]$. 
We now determine the extended twisted TPG for the decomposition given
by (\ref{decom-indecom}). We note that  $V$ can only be connected to two 
nodes corresponding to highest weights
\beq
\nu=\lt\{
\begin{array}{ll}
2\d_1~~({\rm opposite~ parity}), &~~~(c,k)=(a-1,0)\\
2(\d_1+\d_2)~~({\rm same~ parity}), &~~~(c,k)=(a,2).\label{nu-node}
\end{array}
\rt.
\eeq
We thus arrive at the
extended twisted TPG for (\ref{decom-indecom}), given by Figure 1.

\begin{figure}[ht]
\unitlength=1.00mm
\linethickness{0.4pt}
\begin{picture}(129.60,92.00)
\put(47.00,86.00){\line(1,-1){35.00}}
\put(77.00,56.00){\line(-1,0){30.00}}
\put(67.00,51.00){\line(-1,1){20.00}}
\put(47.00,71.00){\line(1,0){15.00}}
\put(47.00,56.00){\line(1,-1){5.00}}
\put(60.00,6.00){\line(1,0){25.00}}
\put(47.00,21.00){\line(1,0){35.00}}
\put(47.00,21.00){\line(1,-1){15.00}}
\put(77.00,6.00){\line(-1,1){20.00}}
\put(127.00,6.00){\line(-1,0){20.00}}
\put(112.00,6.00){\line(-1,1){5.00}}
\put(127.00,6.00){\line(-1,1){20.00}}
\put(112.00,21.00){\line(-1,0){5.00}}
\put(72.00,26.00){\line(1,-1){10.00}}
\put(47.00,86.00){\circle*{5.20}}
\put(47.00,71.00){\circle*{5.20}}
\put(62.00,71.00){\circle*{5.20}}
\put(62.00,56.00){\circle*{5.20}}
\put(47.00,56.00){\circle*{5.20}}
\put(77.00,56.00){\circle*{5.20}}
\put(77.00,21.00){\circle*{5.20}}
\put(62.00,21.00){\circle*{5.20}}
\put(47.00,21.00){\circle*{5.20}}
\put(62.00,6.00){\circle*{3.20}}
\put(62.00,6.00){\circle{5.20}}
\put(77.00,6.00){\circle*{5.20}}
\put(112.00,6.00){\circle*{5.20}}
\put(127.00,6.00){\circle*{5.20}}
\put(112.00,21.00){\circle*{5.20}}
\put(47.00,82.00){\makebox(0,0)[ct]{$+$}}
\put(47.00,67.00){\makebox(0,0)[ct]{$-$}}
\put(47.00,52.00){\makebox(0,0)[ct]{$+$}}
\put(62.00,67.00){\makebox(0,0)[ct]{$-$}}
\put(62.00,52.00){\makebox(0,0)[ct]{$+$}}
\put(77.00,52.00){\makebox(0,0)[ct]{$+$}}
\put(94.00,21.00){\makebox(0,0)[cc]{$\cdots$}}
\put(94.00,6.00){\makebox(0,0)[cc]{$\cdots$}}
\put(38.00,86.00){\makebox(0,0)[rc]{$c=0$}}
\put(38.00,71.00){\makebox(0,0)[rc]{$c=1$}}
\put(38.00,56.00){\makebox(0,0)[rc]{$c=2$}}
\put(38.00,21.00){\makebox(0,0)[rc]{$c=a-1$}}
\put(38.00,6.00){\makebox(0,0)[rc]{$c=a$}}
\put(62.00,80.00){\makebox(0,0)[cb]{$k=1$}}
\put(77.00,65.00){\makebox(0,0)[cb]{$k=2$}}
\put(112.00,29.00){\makebox(0,0)[cb]{$k=a-1$}}
\put(127.00,15.00){\makebox(0,0)[cb]{$k=a$}}
\put(38.00,40.00){\makebox(0,0)[rc]{$\vdots$}}
\put(94.00,39.00){\makebox(0,0)[cc]{$\ddots$}}
\put(47.00,92.00){\makebox(0,0)[cb]{$k=0$}}
\end{picture}
\caption{\small The extended twisted TPG for $U_q[gl(n|n)^{(2)}]~ (n>2)$
for the tensor product $V(\l_a)
\otimes V(\l_a)$. The vertex labelled by the pair $(c,k)$ corresponds to the
irreducible $U_q[osp(n|n)]$ module $V(k, 2a-2c)$ except for the vertex
corresponding to $c=a,~k=1$, which has been circled to indicate that
it is an indecomposable $U_q[osp(n|n)]$-module.
\label{fig3}}
\end{figure}
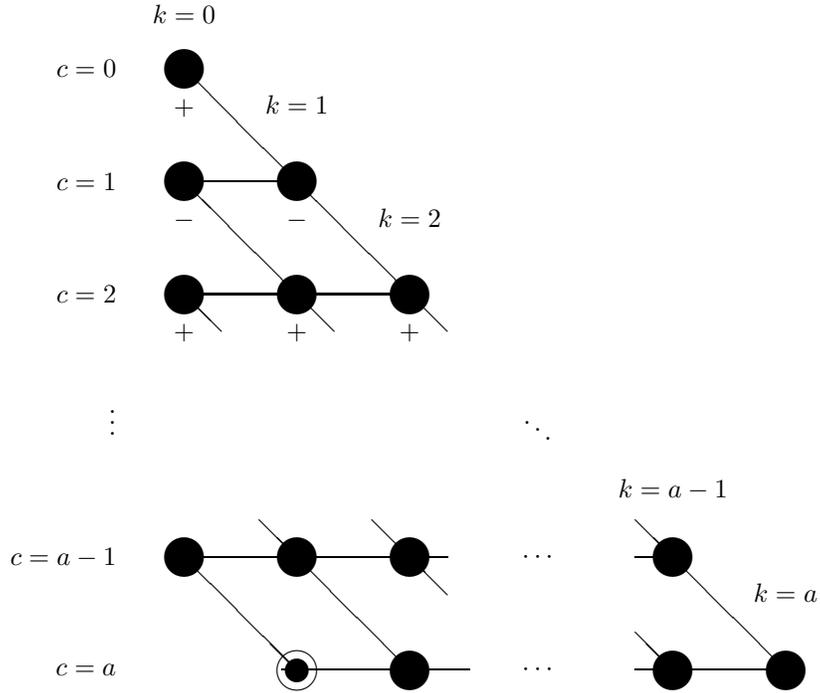

It can be shown that the extended twisted TPG is consistent, i.e. that
the recursion relations (\ref{rec}) give the same result independent of the
path along which one recurses. To prove this it suffices to
show for each closed loop of four vertices in the graph, that the
difference in Casimir eigenvalues for $osp(n|n)$ along one edge equals
the difference along the opposite edge.

Let $P_V\equiv P^{\l_a\l_a}_V$ be the projection operator from 
$V(\l_a)\otimes V(\l_a)$ onto
$V$ and $P_\nu\equiv P^{\l_a\l_a}_\nu$ the projector onto $V(\nu)$.
Then the R-matrix $\check{R}(z)\equiv \check{R}^{\l_a,\l_a}(z)$
from the extended twisted TPG can be expanded in terms of the operators
$N,~P_V$ and $P_\nu$:
\beq
\check{R}(z)=\rho_N(z) N+\rho_V(z) P_V+\sum_\nu\rho_\nu(z) P_\nu.
    \label{r(z)-check}
\eeq
The coefficients $\rho_\nu(z)$ can be obtained recursively
from the extended twisted TPG. However, the coefficients $\rho_N(z)$ and 
$\rho_V(z)$ cannot be read off from the extended twisted TPG since
the corresponding vertex refers to an indecomposable module. 
Rather they are determined by the
approach \cite{Gou97} to $U_q[gl(2|2)^{(2)}]$. The result is
\cite{Gou00}
\bea
\check{R}(z)&=&\rho_N(z) N+\rho_V(z)P_V+{\sum_{c=0}^a}'{\sum_{k=0}^c}' 
  \prod^{c-k}_{j=1} \langle 2j-2a \rangle_+\,\no\\
& &  \prod^c_{i=1} \langle i-2a-1 \rangle_-\,
  P_{(2a-2c+k)\d_1+k\d_2},
\eea
where the primes in the sums signify that terms corresponding to
$c=a$ with $k=0,1$ are ommitted from the sums, and 
$\rho_V(z),~\rho_N(z)$ are given
by 
\bea
\rho_V(z)&=&\frac{z-q^2}{1-zq^2}
   \prod^{a-1}_{j=1} \langle 2j-2a \rangle_+\,
   \prod^{a-1}_{i=1} \langle i-2a-1 \rangle_-,\no\\
\rho_N(z)&=&(-1)^a q^{-a^2}\;\frac{1-z}{1+z}\;\rho_V(z).
\eea

\nonumsection{Acknowledgements}
\noindent
This work has been financially supported by the Australian Research
Council.

\nonumsection{References}

\end{document}